# Ground State of Fermions in a 1D Trap with δ Function Interaction

C. N. Yang

*Tsinghua University, Beijing and Chinese University of Hong Kong, Hong Kong*

*Abstract:  The ground state of Fermions in a 1D trap with δ function interaction is studied mathematically with group theory ideas.*

Marvelous developments in laser cooling and trapping in the last 20 years have created the new field of cold atom research.   One exciting subfield[1] is the physics of trapped one dimensional spin ½ Fermions.   A useful model for such systems is the Hamiltonian:

$$H = \sum_{i=1}^{N}\left[-\frac{\hbar^2}{2m}\frac{\partial^2}{\partial x_i^2} + \frac{1}{2}V(x_i)\right] + g\sum_{i>j}\delta(x_i - x_j). \tag{1}$$

We assume the trapping potential V(x) to approach ∞ when x →—∞ and when x →+∞.   Throughout this paper we consider unharmonic traps as well as harmonic ones.

This Hamiltonian is invariant under the permutation group $S_N$, and contains only space coordinates $x_1, x_2, \ldots x_N$.   Its space eigenfunctions are classified by irreducible representations (IR) of $S_N$. We designate these IRs by their Young graph Y and adopt a simplified notation.   E.g. we write {3,2,1,1} for the graph 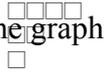 indicating the number of squares in each column.   This is *different* from the standard notation.

By a well known theorem, for Fermions with spin ½ one first forms a spin wavefunction belonging to the IR: Y', where  Y' contains at most 2 rows, and dot product such a spin wavefunction, and its partners under $S_N$, into space wavefunctions belonging to the conjugate IR: Y, with its partners, to form (2J+1) totally antisymmetric wavefunctions  Φ  with total spin equal to one half of the difference of the lengths of the two columns of Y.   Wavefunctions formed in such a manner span the complete Hilbert space of the Fermion system.

<div align="center">CASE g = 0</div>

We now concentrate on the space wavefunctions.   Denote the eigenfunctions for the single particle Hamiltonian by $u_n(x)$, $n = 0, 1, 2, \ldots$ with energies $\varepsilon_0 < \varepsilon_1 < \varepsilon_2 \ldots$ .
When $g = 0$, all eigenfunctions of the full Hamiltonian (1) can be expressed as products of the u's.

This problem had been discussed by Girardeau and Minguzzi[2]. The general problem for $g \neq 0$ is difficult. In an important recent paper Ma, Guan, Chen and Wang found[3] the limit as $g \to +\infty$ of the ground state energy, for each symmetry Y. In this paper we examine the nature of the wavefunction for each symmetry Y, and how it approaches a limit as $g \to +\infty$, by converting the problem to an equivalent problem B which *does not have symmetry requirements*.

To find, when $g = 0$, eigenfunctions belonging to a specific symmetry Y we apply the Young operation QP for that Y, on a product wavefunction of N single particles of the form

$$u_a(x_1)u_b(x_2) \ldots . \tag{2}$$

The operator P is a symmetrizer, and Q an antisymmetrizer. We illustrate this process with the case of $N = 6$, for symmetry Y={3,3}. Starting from the space eigenfunction

$$[u_0(x_1)u_1(x_2)u_2(x_3)][u_0(x_4)u_1(x_5)u_2(x_6)] \tag{3}$$

We obtain

$$\psi_Y = 4 \det[u_0(x_1)u_1(x_2)u_2(x_3)] \times \det[u_0(x_4)u_1(x_5)u_2(x_6)], \tag{4}$$

where we introduce the notation

$$\det[u_0(x_1)u_1(x_2)u_2(x_3)] = \begin{vmatrix} u_0(x_1) & u_1(x_1) & u_2(x_1) \\ u_0(x_2) & u_1(x_2) & u_2(x_2) \\ u_0(x_3) & u_1(x_3) & u_2(x_3) \end{vmatrix}.$$

This $\psi$ is an eigenstate of the Hamiltonian (1) with eigenvalue $(\varepsilon_0 + \varepsilon_1 + \varepsilon_2) + (\varepsilon_0 + \varepsilon_1 + \varepsilon_2)$, belonging to the symmetry {3,3}.

If we had started from another product like (3) where between the 6 indices for the u's there are 3 identical ones, then the resultant ψ is zero because, e.g.

$$\det [u_a(x_1)u_a(x_2)u_6(x_3)] = 0.$$

Thus $\varepsilon_0$ can occur at most twice, also $\varepsilon_1$ can occur at most twice, and we conclude (4) is the *ground state* for Hamiltonian (1).

Wavefunction (4) vanishes on any one of the following 6 planes:

$$x_1 = x_2, \ x_2 = x_3, \ x_3 = x_1, \ x_4 = x_5, \ x_5 = x_6 \ \text{ and } x_6 = x_4. \tag{5}$$

Now consider the open region $R_Y$ defined by

$$R_Y: \quad x_1 < x_2 < x_3 \quad \text{and} \ x_4 < x_5 < x_6. \tag{6}$$

Figure 1 shows that the region $x_1 < x_2 < x_3$ is bounded by the two planes $x_1 = x_2$ and $x_2 = x_3$. Thus $R_Y$ is bounded by the 4 planes

$$x_1 = x_2, \ x_2 = x_3, \ x_4 = x_5, \text{ and } x_5 = x_6. \tag{7}$$

It is important to notice that $\psi_Y$ vanishes on the boundary planes (7) of $R_Y$. Figure 1 also shows that altogether there are 6×6 = 36 regions like $R_Y$ defined by (6).

## AN EQUIVALENT PROBLEM

We state and prove for space wavefunctions:

Theorem 1: For any value of g, consider 2 different eigenvalue problems:
(A) Hamiltonian H with symmetry Y={3,3} in full $\infty^6$ space, and
(B) Hamiltonian H in region $R_Y$ with the boundary condition that the wavefunction vanishes on its surface (7). [Notice this boundary condition is Y dependent].
The eigenvalues of the two problems are identical, and the corresponding unnormalized wavefunctions are proportional in region $R_Y$.

Proof: Starting from a solution of problem B, by analytic continuation beyond the 4 boundary planes (7), one obtains the space eigenfuction in full $\infty^6$ space.
Conversely starting from a solution ψ of problem A, the wavefunction QPψ vanishes on the boundary planes (7) of $R_Y$. Thus if it is nonvanishing, it is an eigenfunction for problem B.     QED

[Notice this theorem is the generalization of the simple theorem for 2 particles: Any antisymmetric wavefunctions in full $\infty^2$ space can be continued from a wavefunction in half space that vanishes on the boundary $x_1 = x_2$.]

Theorem 2:    For any value of g, the ground state wavefunction for problem B has no zeros in the interior of $R_Y$, and is not degenerate.

Proof:    This is a special case of the general theorem that the ground state wavefunction has no zeros in the interior, if we do not impose symmetry conditions.    QED

It follows from these two theorems that for symmetry {3,3}, the ground state of Hamiltonian (1) is not degenerate.

$$\text{CASE } g \neq 0$$

For $g \neq 0$, we take advantage of Theorem 1 and study eigenvalue problem B instead of the original eigenvalue problem. The advantage of B is that no symmetry condition is imposed.    Starting from wavefunction (4) at $g = 0$, we follow $\psi_Y$ as g changes. According to theorem 2, it is everywhere $\geq 0$ in $R_Y$.    It vanishes on the 4 boundary planes (7) of $R_Y$.    Continuing them into the full $\infty^6$ space gives an eigenfunction of problem A with Y={3,3}.

There are altogether 15 delta function interactions in (1).    Four of them reside on boundary (7) of $R_Y$.    Figure 1 shows that two of them, $\delta(x_1 - x_3)$ and $\delta(x_4 - x_6)$ are entirely outside of $R_Y$.    The remaining 9 reside on nine planes.

$$x_i - x_j = 0, \quad i = 1, 2 \text{ or } 3, \quad j = 4, 5, \text{ or } 6 \qquad (8)$$

each of which partly lies inside region $R_Y$.    Inside $R_Y$, when g increases, *V shaped cusps* are formed on wavefunction $\psi_Y$ at these nine planes, depressing its value on the plane.    But by Theorem 2, $\psi_Y$ remains $> 0$ in the interior of region $R_Y$.    When g $\to +\infty$, the cusps become infinitely deep.    I.e. $\psi_Y \to 0$ on the nine planes inside $R_Y$.    We denote this limiting wavefunction by $\psi_{Y\infty}$.    In $R_Y$ it vanishes only on the nine planes (8) and on the boundary of $R_Y$ [the 4 planes (7).].

Now the nine planes (8) divide $R_Y$ into $\frac{6!}{3!3!} = 20$ subregions, such as

$$x_1 < x_4 < x_2 < x_5 < x_3 < x_6 \quad \text{or} \quad x_4 < x_5 < x_1 < x_2 < x_3 < x_6,$$

each of which conforms with the conditions $x_1 < x_2 < x_3$ and $x_4 < x_5 < x_6$.

Consider now the totally antisymmetric space eigenfunction $\psi_{AS}$ of the Hamiltaonian (1) with energy $\varepsilon_0+\varepsilon_1+\varepsilon_2+\varepsilon_3+\varepsilon_4+\varepsilon_5$. It belongs to $\{6,0\}$, and is the ground state with such symmetry. In the interior of each subregion it is either all $> 0$ or all $< 0$.

Now we compare $\psi_{AS}$ with $\psi_{Y\infty}$ in any subregion of $R_Y$. They are both eigenfunctions of H. They both vanish on the boundary of the subregion. They both have no zeros in the interior of this subregion. Thus in each subregion of $R_Y$ they must be proportional to each other, and *have the same energy* $\varepsilon_0+\varepsilon_1+\varepsilon_2+\varepsilon_3+\varepsilon_4+\varepsilon_5$.

Proceeding this way we see easily that

$$\psi_{Y\infty} = |\psi_{AS}| \quad \text{in } R_Y.$$

This is a generalization of the early result of Girardeau[4]. Continuing $\psi_{Y\infty}$ into full $\infty^6$ space we have

$$|\psi_{Y\infty}| = |\psi_{AS}|. \tag{9}$$

$$Y = \{4,2\}$$

We now consider $Y = \{4,2\}$ and start with the product

$$[u_0(x_1)u_1(x_2)u_2(x_3)\,u_3(x_4)][u_0(x_5)u_1(x_6)]$$

arriving at the eigenfuction for $g = 0$:

$$\psi_Y = \det[u_0(x_1)u_1(x_2)u_2(x_3)u_3(x_4)] \times \det[u_0(x_5)u_1(x_6)] \tag{10}$$

with energy $(\varepsilon_0+\varepsilon_1+\varepsilon_2+\varepsilon_3) + (\varepsilon_0+\varepsilon_1)$.

This $\psi_Y$ vanishes on the boundary of region $R_Y$

$$R_Y : \quad x_1 < x_2 < x_3 < x_4 \quad \text{and} \quad x_5 < x_6 \tag{11}$$

bounded by four planes. [Notice this $R_Y$ is different from the $R_Y$ for symmetry $\{3,3\}$.] Repeating the arguments for the case $Y = \{3,3\}$ we arrive at the conclusion that as $g \to \infty$, $\psi_Y$ approach a limit $\psi_{Y\infty}$, with $E \to \varepsilon_0+\varepsilon_1+\varepsilon_2+\varepsilon_3+\varepsilon_4+\varepsilon_5$. Continuation across the boundary planes extends $\psi_{Y\infty}$ into full $\infty^6$ space, and

$$|\psi_{Y\infty}| = |\psi_{AS}|.$$

Thus we have

Theorem 3: The ground state of Hamiltonian (1) for symmetry Y is not degenerate. Denote by $E_Y$, its energy. Then

at $g = 0$ 
$$E_{\{3,3\}} = (\varepsilon_0+\varepsilon_1+\varepsilon_2) + (\varepsilon_0+\varepsilon_1+\varepsilon_2) \quad , \quad (J=0),$$
$$E_{\{4,2\}} = (\varepsilon_0+\varepsilon_1+\varepsilon_2+\varepsilon_3) + (\varepsilon_0+\varepsilon_1) \quad , \quad (J=1),$$
$$E_{\{5,1\}} = (\varepsilon_0+\varepsilon_1+\varepsilon_2+\varepsilon_3+\varepsilon_4) + \varepsilon_0 \quad , \quad (J=2),$$
$$E_{\{6,0\}} = \varepsilon_0+\varepsilon_1+\varepsilon_2+\varepsilon_3+\varepsilon_4+\varepsilon_5 \quad , \quad (J=3).$$

Furthermore $E_{\{6,0\}}$ is independent of g. As g increases the other three all increase *monotonically* and approach $E_{\{6,0\}}$ as $g \to +\infty$.

To summarize, for the ground state wavefunction at $g = 0$, we choose for each Y, such functions as (4) and (9) which are products of two determinants. Then we follow these wavefunctions as $g \to +\infty$. They approach limits $\psi_{Y\infty}$ which satisfy equation (9), each vanishing on all 15 planes defined by the delta function interactions. Each $\psi_{Y\infty}$ is positive in one half of the $6!=720$ subregions, and negative in the other half. For different Y's $\psi_{Y\infty}$ are orthogonal to each other. Notice they are *cusp-less on different sets of planes for different Y's*.

Furthermore according to a theorem due to Lieb and Mattis[5], for all finite g values

$$E_{\{3,3\}} < E_{\{4,2\}} < E_{\{5,1\}} < E_{\{6,0\}}.$$

Obviously these results can be generalized to any even or odd values of N.

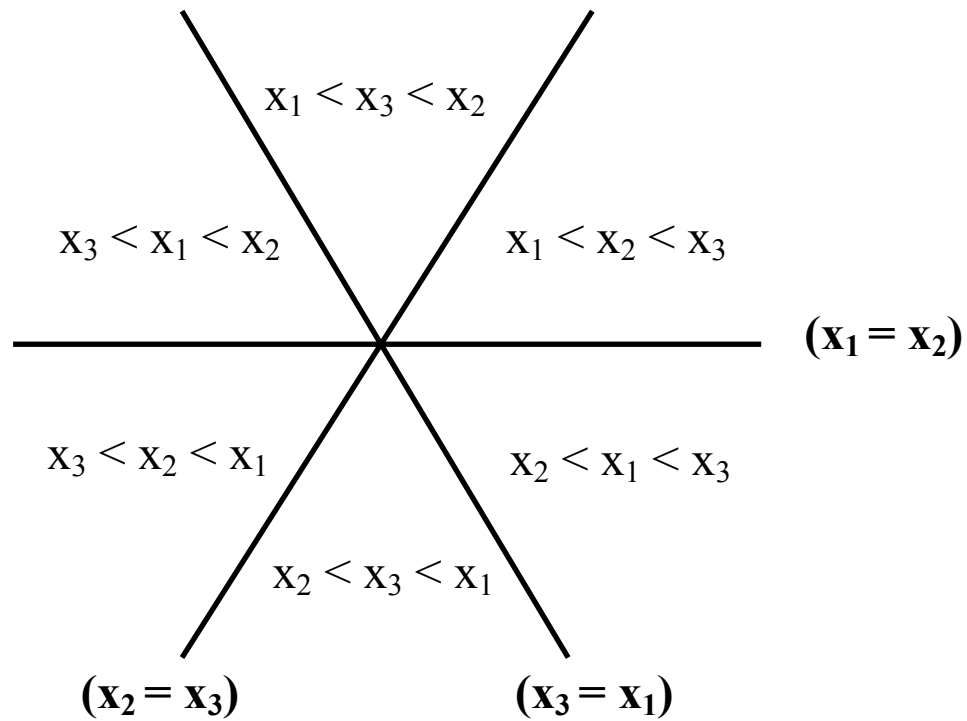

Figure 1.   The 3 planes $x_1 = x_2$, $x_2 = x_3$ and $x_3 = x_1$, divide $\infty^3$ space into 6 regions as indicated.